\documentclass[11pt]{article}
\newcommand{\bc}{\begin{center}}
\newcommand{\ec}{\end{center}}
\newcommand{\be}{\begin{equation}}
\newcommand{\ee}{\end{equation}}
\newcommand{\ba}{\begin{array}}
\newcommand{\ea}{\end{array}}
\newcommand{\bea}{\begin{eqnarray}}
\newcommand{\eea}{\end{eqnarray}}
\newcommand{\edc}{\end{document}}

\usepackage{graphics}
\usepackage{amsmath,amsxtra,amssymb,latexsym, amscd,color}
\usepackage[mathscr]{eucal}
\usepackage{graphicx}
\begin{document}
\thispagestyle{empty}

\begin{center}

\vspace{2.5cm}
{\bf FUNCTIONAL INTEGRATION AND HIGH ENERGY SCATTERING OF PARTICLES WITH ANOMALOUS MAGNETIC MOMENTS IN QUANTUM FIELD THEORY  }\\

\vspace{0.5cm}

 Nguyen Suan Han $^{a,b,}$\footnote{Email: lienbat76@gmail.com}, Le Hai Yen $^{a}$,  Nguyen Nhu Xuan $^{c}$\\
 \vspace{0.5cm}
{$^a$\it Department of Theoretical Physics,Vietnam National University, Hanoi, Vietnam.} \\
{$^b$\it The Abdus Salam International Centre for Theoretical Physics, Trieste, Italy.}\\
{$^c$\it Department of Theoretical Physics, Le Qui Don Technical University, Hanoi, Vietnam.}\\%

\end{center}

\vspace{0.5cm}
\centerline{\bf Abstract}
\baselineskip=18pt
\bigskip

The functional integration method is used for studying the scattering of a scalar pion on nucleon with the anomalous magnetic moment in the framework of nonrenomalizable quantum field theory. In the asymptotic region $s\rightarrow\infty $, $|t|\ll s $  the representation of eikonal type for the amplitude of pion-nucleon scattering is obtained. The anomalous magnetic moment leads to additional terms in the amplitude which describe the spin flips in the scattering process. It is shown that the renormalization problem does not arise in the asymptotic $s \rightarrow\infty$ since the unrenomalized divergences disappear in this approximation. Coulomb interference is considered as an application.\\

\textbf{ Keywords:} Quantum scattering; anomalous magnetic moment.\\


\newpage

\section{Introduction}

The eikonal approximation for the scattering amplitude of high-energy particles in quantum field theory including quantum gravity has been investigated by many authors using various approaches $[1-17]$. Nevertheless, these investigations do not take into account the spin structure of the scattering particles. It is, however, well known from recent experiments that spin effects are important in many processes $[18-20]$. This motivates us to study the problem of generalizing the functional integration method allowing for the spin effects; namely, we consider the scattering of particles with anomalous moments.\\

Here, we investigate the electromagnetic interaction, i.e., the interaction due to the exchange of vector particles with vanishing mass $\mu\rightarrow 0$. It is pointed out that the eikonal approximation works well in a wide energy range $[21-23]$. This approximation was applied to the problem of bound states, not only the Balmer formula was obtained but also the relativistic corrections to the ground level energy $[5,24]$.\\

The interaction between a particle with an anomalous magnetic moment and an electromagnetic field is nonrenormalizable $[25,26]$. Since ordinary perturbation theory does not work in nonrenormalizable field theories $[27-29]$, in this work we use the functional integration which enables us to perform the calculations in a compact form.\\

The rest of this article is organized as the following. In the second section, we consider the scattering of a scalar pion on a nucleon with an anomalous magnetic moment. Using the exact expression of the single-particle Green's function in the form of a functional integral, we obtain the two-particle Green's function by the averaging of two single-particle Green's function. By transition to the mass shell of external two-particle Green's function, we obtain a closed representation for the $ \pi N$  elastic scattering amplitude expressed in the form of the functional integrals. To estimate the functional integrals we use the straight line path approximation, based on the idea of rectilinear paths of interacting particles of asymptotically high energies and small momentum transfers. The third section is devoted to investigating the asymptotic behavior of this amplitude in the limit of high energies $s\rightarrow\infty $, $|t|\ll s $, and we obtained an eikonal or Glauber representation of the scattering amplitude. As an application of the eikonal formula obtained in fourth section, we consider the Coulomb interference in the scattering of charged hadrons. Here, we find a formula for the phase difference; this is a generalization of the Bether formula in the framework of relativistic quantum field theory. Finally, concluding remarks are presented.\\

\section{Construction of the two-particle scattering amplitude}

We consider the scattering of a scalar particle (pion $\pi$) on a Dirac particle with anomalous magnetic moment (nucleon $N$)\footnote{For simplicity, pion will be regarded as particle 1 and nucleon as particle 2.} at high energies and at fixed transfers in quantum field theory. To construct the representation of the scattering in the framework of the functional approach we first find the two-particle Green's function, once the Green's function is obtained we consider the mass respective to the external ends of the two particle lines. \\

Using the method of variational derivatives we shall determine the two particles Green's function $G_{12}(p_1,p_2 | q_1,q_2)$ by the following formula:
$$ G_{12}(p_1,p_2 | q_1,q_2)=\exp \Bigg [\frac{i}{2}\int d^{4}k D_{\mu\nu}(k)\frac{\delta^{2}}{\delta A_{\mu}(k)\delta A_{\nu}(-k)}\Bigg]
G_{1}\left(p_{1},q_{1}|A\right)G_{2}\left(p_{2} ,q_{2}|A\right).S_{0}(A)\Bigr|_{A=0}, \eqno(2.1)$$
where $S_{0}(A)$ is the vacuum expectation of the $S$ matrix in the given external field $ A $. For simplicity, we shall henceforth ignore vacuum polarization effects and also the contributions of diagrams containing closed nucleon loops; $G_{1}(p_1,q_1|A)$-the Fourier of the Green's function $(A.5)$ of particle $1$ in the given external field takes the form
$$
G_{1}(p_1,q_1|A)=i\int_{0}^{s}ds e^{i(p_{1}^2-m_{1}^2)s}\int d^{4}x e^{i(p_1-q_1)x}\int[\delta^{4}\nu]_{0}^{s}\exp[ie\int_{0}^{s}J_{\mu}A_{\mu}], \eqno(2.2)
$$
here we use the notation $ \int JA=\int J_{\mu}(z)A_{\mu}(z)$, and $J_{\mu}(z)$ is the current of the particle 1  defined by
$$
J_{\mu}(z)=2\int_{0}^{s}d\xi \nu_{\mu}(\xi)\delta(z-x_i +2\int_{0}^{\xi} [\nu_i(\eta)+p_i] d\eta). \eqno(2.3)
$$
We notice that on the mass shell the ordinary Green's function $ G_{2}(p_2,q_2|A)$ and the squared Green's functions $\overline{G}_{2}(p,q|A) $ are identical $[4]$, in eq. (2.1) we thus use the latter in eq. $(A.11)$ (appendix)
$$
\overline{G}_{2}(p_2,q_2|A)=i\int_{0}^{s}e^{i(p_{2}^2-m_{2}^2)s}ds \int d^{4}x e^{i(p_2-q_2)x} T_{\gamma}\int[\delta^{4}\nu]_{0}^{s} \exp\{ie\int_{0}^{s}J_{\mu}A_{\mu}(x)\}, \eqno(2.4)
$$
where $ T_\gamma$ is the symbol of ordering the $\gamma_{\mu}$ matrices  with respect to the ordering index $\xi$, and $J_{\mu}(z)$ is the current of particle 2 defined by
$$
J_{\mu}(z)=2\int_{0}^{s} d\xi[\nu_{\mu}(\xi)+ \frac{1}{2}\sigma_{\mu\nu}(\xi)i\partial_{\nu}]\delta(z-x_i +2\int_{0}^{\xi}[\nu_i(\eta)+p]d\eta). \eqno(2.5).
$$
Substituting $(2.2), (2.4)$ into $(2.1)$ and performing variational derivatives, for the two-particle Green's function we find the following expression:
$$
G_{12}(p_1,p_2|q_1,q_2)=i^2 \prod_{i=1,2}\Bigg (\int_{0}^{\infty}ds_i e^{i(p_{i}^2-m_{i}^2)s_i} \int[\delta^{4}\nu_i]_{0}^{s_i}\int d^{4}x_i e^{i(p_i-q_i)x_i} \exp [-\frac{ie^2}{2}\int D(J_{1}+J_{2})^2] \Bigg ), \eqno(2.6)
$$
where we introduce the abbreviated notion
$$
JDJ=\int dz_1 dz_2 J_{\mu}(z_1)D_{\mu\nu}(z_1-z_2)J_{\nu}(z_2).
$$
Expanding expression $(2.6)$ with respect to the coupling constant $ e^2 $ and taking the functional integrals with respect to $\nu_i (\eta)$, we obtain the well-known series of perturbation theory for the two-particle Green's function.
The term in exponent $(2.6)$ we can rewrite in the following form:
$$
-\frac{ie^2}{2}\int D(J_1+J_2)^2=-ie^2 \int DJ_1J_2 -\frac{ie^2}{2}\int DJ_{1}^{2}-\frac{ie^2}{2}\int DJ_{1}^{2}, \eqno(2.7)
$$
the first term on the right-hand side $(2.7)$ corresponds to the one-photon exchange between the two-particle and the remainder lead to radiative corrections to the lines of the two-particles.\\

The scattering amplitude of two particles is expressed in the two particles Green's function  by equation:
$$
i(2\pi)^{4}\delta ^{(4)} \left(p_{1} +p_{2} -q_{1}
-q_{2}\right)T\left(p_{1},p_{2} |q_{1} ,q_{2}\right)=$$
$$
=\frac{1}{2m_{2}} \overline{u}(q_2)\Bigg [lim_{p_{i}^{2},q_{i}^{2}\rightarrow m_{i}^{2}}(p_{i}^{2}-m_{i}^{2})G_{12}(p_1,p_2 | q_1,q_2) (q_{i}^{2}-m_{i}^{2}) \Bigg]u(p_2), \eqno(2.8)
$$
the spinors $\overline{u}(q_2)$ and $ u(p_2)$ on the mass shell satisfy the Dirac equation and the normalization condition $\overline{u}(q_2)u(p_2)=2m_2$. \\

The transition to the mass shell $ p_{i}^{2}; q_{i}^{2} \rightarrow m_{i}^{2} ; $ calls for separating from formula $(2.8)$ the pole terms $ (p_{i}^{2}-m_{i}^{2})^{-1}$ and $(q_{i}^{2}-m_{i}^{2})^{-1}$ which cancel the factors $ (p_{i}^{2}-m_{i}^{2})$ and $(q_{i}^{2}-m_{i}^{2})$. In perturbation theory this compensation is obvious, since the Green's function is sought by means of methods other than perturbation theory, the separation of the terms entails certain difficulties. We shall be interested in the structure of scattering amplitude as a whole, therefore the development of a correct procedure for going to the mass shell in the general case is very important. Many approximate methods that are reasonable from the physical point of view when used before the transition on the mass shell , shift the positions of the pole of the Green's function and render the procedure of finding the scattering amplitude mathematically incorrect. In present paper we shall use a method for separating the poles of the Green's functions that generalizes the method introduced in Ref. $[30]$ to finding the scattering amplitude in a model of scalar nucleon interacting with scalar meson field, in which the contributions of closed nucleon loops are ignored.\\

Substituting $(2.6)$ into $(2.8)$ we get

$$
(2\pi)^{4}\delta^{4}(p_1+p_2-q_1-q_2)iT(p_1,p_2|q_1,q_2)$$
$$
=\frac{1}{2m_{2}} \overline{u}(q_2)\Bigg [  lim_{p_{i}^{2}, q_{i}^{2}\rightarrow m_{i}^{2}} \Bigg (\prod_{i=1,2}(p_{i}^{2}-m_{i}^{2})(q_{i}^{2}-m_{i}^{2}) \int d^{4}x_{i} e^{i(p_i-q_i)x_i}\int_{0}^{\infty}ds_i \int_{0}^{\infty} d\xi e^{i(p_{i}^{2}-m_{i}^{2} )} \Bigg ) $$
$$e^2 DJ_{1}J_{2}\int_{0}^{1} d\lambda \exp [-ie^2\lambda\int DJ_{1}J_{2} ] \Bigg]u(p_2). \eqno(2.9)
$$

Deriving $(2.9)$ we employ the operator of subtracting unity in the formula $(2.9)$ from the exponent function containing the D-function in its argument in accordance with

$$
e^{-ie^2 \int DJ_{1}J_{2}}-1=-ie^2 \int_{0}^{1} d\lambda DJ_{1}J_{2} e^{-i\lambda \int DJ_{1}J_{2}}.
$$

This corresponds to eliminating from the Green's function the terms describing the propagation of two noninteracting particles. Taking into account the identity:
$$
\prod_{k=1,2}\int_{0}^{\infty} ds_{k} \int_{0}^{s_k} d\xi_{k}...\rightarrow \prod_{k=1,2}\int_{0}^{\infty} d\xi_{k} \int_{\xi}^{\infty} ds_{k}...
$$
and making a change of the ordinary and the functional variables

$$
s_i\rightarrow s_i +\xi_i;   i=1,2,
$$
$$
x_i\rightarrow x_i-2\int_{0}^{\xi_{i}} [p+\nu(\eta)]d\eta,
$$
$$
\nu_{i}(\eta)\rightarrow \nu_{i}(\eta-\xi)-(p-q)\theta(\eta-s_i),
$$
we transform eq. $(2.9)$ into
$$
(2\pi)^{4}\delta^{4}(p_1+p_2-q_1-q_2)iT(p_1,p_2|q_1,q_2)=$$
$$
=\frac{1}{2m_{2}} \overline{u}(q_2)\Bigg [  lim_{p_{i}^{2}, q_{i}^{2}\rightarrow m_{i}^{2}} \Bigg (\prod_{i=1,2}(p_{i}^{2}-m_{i}^{2})(q_{i}^{2}-m_{i}^{2}) \int d^{4}x_{i} e^{i(p_i-q_i)x_i}$$
$$
\int_{0}^{\infty}d\xi_{i}e^{i(p_{i}^{2}-m_{i}^{2})\xi} \int_{0}^{\infty}ds_{i}e^{i(q_{i}^{2}-m_{i}^{2})s_i}\Bigg)
\int[\delta^{4}\nu_{1}]_{\xi_1}^{s_1}\int[\delta^{4}\nu_{2}]_{\xi_2}^{s_2}$$
$$e^2 DJ_{1}J_{2}\int_{0}^{1} d\lambda \exp [-ie^2\lambda\int DJ_{1}J_{2} ] \Bigg]u(p_2). \eqno(2.10)
$$
In the following we consider the forward scattering, and the radiative corrections to lines of the particles in eq $(2.10)$ will be omitted.
We now note that the integrals with respect to $ s_i $ and $\xi_i $ give factors $(p_{i}^{2}-m_{i}^{2})^{-1}$ and $(q_{i}^{2}-m_{i}^{2})^{-1}$; $i=1,2 $. Therefore, in eq. $(2.10)$ we can go over the mass shell with respect to the external lines of the particle  using the relations $[31]$

$$
lim_{a,\varepsilon \rightarrow 0}\Bigg (ia\int_{0}^{\infty} e^{ias-\epsilon} f(s)\Bigg) =f(\infty),
$$
which holds for any finite function $f(s)$. By means of the substitutions $ x_1=(y+x)/2$ and $x_2=(y-x)/2$ in eq. $(2.10)$ and performing the integration with respect to $ dy $ we can separate out the $\delta$-function of the conservation of the four-momentum $\delta^{4}(p_1+p_2-q_1-q_2)$. As a result, the scattering amplitude takes the form

$$
T(p_1,p_2|q_1,q_2)=\frac{1}{2m_{2}} \overline{u}(q_2)\Bigg [ e^2 \prod_{i=1,2}\int[\delta^{4}\nu_{i}]_{-\infty}^{\infty}\int d^{4}x e^{i(p_1-q_1)x}\times$$
$$[p_1+q_1+2\nu (0)]D^{\mu\nu}(x)[p_2+q_2+2\nu(0)]_{\nu}\int_{0}^{1}d\lambda \exp [-ie^2\lambda\int DJ_{1}J_{2}] \Bigg]u(p_2), \eqno(2.11)$$
where
$$
J_{1\mu}(k;p_1,q_1|\nu)=2\int_{-\infty}^{\infty}d\xi [p_1\theta(\xi)+q_1\theta(-\xi)+\nu(\eta)]_{\mu} \exp\Big \{ 2ik[p_{1}\xi\theta(\xi)+q_{1}\xi\theta(-\xi) +\int_{0}^{\xi}\nu_{1}(\eta)d\eta] \Big\},
$$
$$
J_{2\mu}(k;p_2,q_2|\nu)=2\int_{-\infty}^{\infty}d\xi \Big \{[p_2\theta(\xi)+q_2\theta(-\xi)+\nu(\eta)]_{\mu}+\frac{1}{2}\sigma_{\mu\nu}(\xi)i\partial_{\nu} \Big \} \times$$
$$ \exp\Big \{ 2ik[p_{2}\xi\theta(\xi)+q_{2}\xi\theta(-\xi) +\int_{0}^{\xi}\nu_{2}(\eta)d\eta] \Big\}.
$$

Here, $\exp[-ie^2\lambda\int DJJ]$ describes virtual-photon exchange among the scattering particles. The integration with respect to $ d\lambda$
ensures subtraction of the contribution of the freely propagating particles from the matrix element. By going over to mass shell of external two particle Green's function, we obtain an exact closed representation for the "pion-nucleon" elastic scattering amplitude, expressed in the form of the double functional integrals. We would like to emphasize that eq. $(2.11)$ can be applied for different ranges of energy.

\section{Asymptotic behavior of the scattering amplitude at high energy}

The important point in our method is that the functional integrals with respect to $\delta^{4}\nu$ are calculated by the straight-line path approximation $[2,3]$, which corresponds to neglecting the functional variables in the arguments of the $D$-functions in $(2.11)$. In the language of Feynman diagrams, this linearizes the particle propagators with respect to the momenta of the virtual photon. Therefore, the scattering amplitude $(2.11)$ in this approximation takes the form\\
$$
T(p_1,p_2|q_1,q_2)=$$
$$\frac{1}{2m_{2}} \overline{u}(q_2)\Bigg [ e^2 \int d^{4}x e^{i(p_1-q_1)x}
[p_1+q_1]D^{\mu\nu}(x)[p_2+q_2]_{\nu}\int_{0}^{1}d\lambda \exp [-ie^2\lambda\int DJ_{1}J_{2}] \Bigg]u(p_2). \eqno(3.1)$$

We perform the following calculation in the center -of-mass system of colliding particles $\overrightarrow{p}_1=-\overrightarrow{p}_2=\overrightarrow{p}$  and we direct the z-axis along the momentum $\overrightarrow{p}_1$:
 $$ p_1=(p_{10},0,0, p=p_z); p_2=(p_{20},0,0,-p),$$
$$
s=(p_{10}+p_{20})^2=4p_{0}^2; p_{10}=p_{20}=p_{0},   t=(p_1-q_1)^2=(p_2-q_2)^2; \eqno(3.2)
$$
integrating over $db_0$ and $db_z$ in $(3.1)$ we obtain for the scattering amplitude
$$T(s,t)=-2is\frac{\overline{u}(q_2)}{2m_2}\int d\overrightarrow{b}_{\bot}e^{i\Delta\overrightarrow{b}_{\bot}} \times$$
$$\times\Bigg (T_{\gamma}\exp\{ie\int_{-\infty}^{\infty}d\tau_{1}\int_{-\infty}^{\infty}d\tau_{2} J_{1\mu}(\widehat{p_1}^{\mu})D_{\mu\nu}(b_{\tau_{1}\tau_{2}})J_{2\nu}(\widehat{p}_{2},\gamma(\tau_2)) \} -1 \Bigg)u(p_2), \eqno(3.3)
$$
where $\widehat{p}_{i}^{\mu}=p_{i}^{\mu}/|p|$, $\tau_{i}=2|p|\xi_{i}$, $(i=1,2)$, $b_{\tau_{1}\tau_{2}}=\overrightarrow{b}\perp-p_{1}\tau_{1}+p_{2}\tau_{2}$.

Let us consider the asymptotic behavior of the elastic forward amplitude of the two-particles $(3.1)$ in the region $s\rightarrow\infty $, $|t|\ll s $.  In this region, spinors $u(p)$ and $\overline{u}(p)$, which are solutions of the Dirac equation $[25]$

$$ u(p)=\left(\begin{array}{c} {1}
\\ {\frac{\vec{\sigma }\vec{p}}{|p|} }
\end{array}\right)\sqrt{m} \psi _{p},\bar{u}(q)=\bar{\psi}_q\sqrt{m} \left(1,\frac{\vec{\sigma }\vec{p}}{|p|}
\right),\quad |\vec{p}|\approx |\vec{q}|,  \eqno(3.4)
$$
where $\psi_p$ and $\overline{\psi}_q $ are ordinary two-component spinors. \\

Using the expansion of $J^\mu [\widehat{p}_2, \gamma (\tau_2)]$ with respect to the $z$ component of the momentum  and substituting $(3.4)$ into $(3.3)$, we obtain
$$
T(s,t)=-2is\bar{\psi}_{q_2} \int d\vec{b}_\bot e^{i\Delta
\vec{b}_\bot} \left[e^{i\chi_0(b)} \Gamma _1(b)-1\right]\psi _{p_2}, \eqno(3.5)
$$
where $\chi_0(b)$ is the phase corresponding to the Coulomb interaction. This phase is determined by
$$
\chi _0(b)=\frac{e^2}{(2\pi)^2} \int
d\vec{k}_\bot\frac{e^{-i\vec{k}_\bot\vec{b}_\bot}}{\mu ^2
+\vec{k}_\bot^2} =\frac{e^2}{2\pi}
K_0\left(\mu|\left|\vec{b}_\bot\right|\right),\eqno(3.6)
$$
where $ K_0\left(\mu \left|\vec{b}_\bot\right|\right)$- is the MacDonald function of zeroth order, and the expression $ \Gamma _1 (b)$
is equal to
$$
\Gamma _1(b)=\frac{1}{2}(1,-\sigma _z)T_{\tau
_2}\exp\Biggr\{-i\kappa\int _{-\infty }^\infty d\tau _1\int
_{-\infty}^\infty d\tau _2\Bigr[\hat{p}_1^\mu\vec{\gamma}^\bot(\tau
_2)\times \vec{\partial}_\bot D_{\mu \rho }^c\left(b_{\tau _1 \tau
_2}\right)\hat{p}_2^\rho- $$
$$-\hat{p}_1^\mu\left[\gamma^z(\tau _2)+\gamma ^0(\tau _1)\frac{p_z}{p_0}\right]\times
\left[\partial_z D_{\mu 0}^c\left(b_{\tau _1\tau _2}\right)-\partial
_0 D_{\mu z}^c\left(b_{\tau _1 \tau
_2}\right)\right]\Biggr\}\left(\begin{array}{c} {1}\\ {-\sigma _z}
\end{array}\right). \eqno(3.7)
 $$
Note that the expansion of the last expression in a series in powers of $\left[\gamma^z+\gamma^0\frac{p_z}{p_0}\right]$ is actually with
respect to $\left[\gamma^z +\gamma^0 \frac{p_z}{p_0}\right]^2=-\frac{m^2}{p_0^2}$, since  $(1,-\sigma_z)\left[\gamma^z +\gamma^0 \frac{p_z}{p_0}\right] (1,-\sigma_z )=0$. Therefore, the second term in the argument of the exponent in $(3.7)$ can be ignored altogether. Thus, we have

$$\Gamma _1(b)=\frac{1}{2} \left(1,\sigma _z\right)T_{\tau
_2}\exp\left[-2e\kappa \int_{-\infty }^\infty d\tau _1\int
_{-\infty}^\infty d\tau _2\vec{\gamma }_\bot(\tau _2)\vec{\partial
}_\bot D\left(b_{\tau _1 \tau
_2}\right)\right]\left(\begin{array}{c} {1}\\ {-\sigma _z}
\end{array}\right). \eqno(3.8)
 $$
Since
 $$
\left[\vec{\gamma }_\bot (\tau _2)\vec{\partial}_\bot D_0^c
\left(b_{\tau _1 \tau _2}\right),\vec{\gamma }_\bot(\tau
'_2)\vec{\partial }_\bot D_0^c \left(b_{\tau _1 \tau '_2}
\right)\right]\Bigr|_{\tau '_2 \neq\tau _2}, \eqno (3.9)
 $$
the $\vec{\gamma}_\bot(\tau_2)$ matrix in $(3.8)$ does not depend on the ordering parameter $ \tau_2 $ and the $T_{\tau_2} $ ordering
exponential is equal to the ordinary exponential :
 $$
\Gamma _1(b)=\frac{1}{2} \left(1,\sigma _z\right)\exp
\left[-2e\kappa\vec{\gamma}_\bot\vec{\partial}_\bot\int _{-\infty
}^\infty d\tau _1\int _{-\infty }^\infty d\tau _2 D\left(b_{\tau
_1 \tau _2}\right)\right]\left(\begin{array}{c} {1} \\
{-\sigma _z}
\end{array}\right)$$
$$
=\frac{1}{2} \left(1,\sigma _z\right)\exp \left[-\frac{e\kappa
}{2\pi} \vec{\gamma }_\bot\vec{\partial }_\bot K_0\left(\mu
|\vec{b}_\bot\right)\right]\left(\begin{array}{c} {1}
\\ {-\sigma _z}\end{array}\right)
\eqno (3.10)
 $$
We go over to cylindrical coordinates $ \vec{b}_\bot=\vec{\rho}=\rho \vec{n}$, $\vec{n}=(cos\phi, sin\phi)$, $\phi$ is the azimuthal
angle in the plane $(x,y)$. Remembering further that
$$
\left[\vec{n}\times\vec{\sigma}\right]_z =-\sigma _x \sin \varphi
+\sigma\cos \varphi, \quad \left[\vec{n}\times
\vec{\sigma}\right]_z^2 =1,\eqno(3.11)
 $$
we obtain
$$\Gamma _1(b)=\exp \left\{i\left[\vec{n}\times \vec{\sigma
}\right]_z\chi _1(\vec{\rho})\right\}\eqno(3.12)$$ where $
\chi_1(\rho)$ is determined by
$$
\chi _1(\vec{\rho})=\frac{e\kappa}{2\pi}\partial _\rho K_0 \left(\mu
|\rho|\right) \eqno(3.13)
 $$
As a result , we obtain the eikonal representation for the $\pi N$ scattering amplitude\footnote{Scattering amplitude $T(s,t)$ in c.m.s
can be normalized by the expression $\sigma _{tot}=\frac{ImT (s,t=0)}{s},\quad \frac{d\sigma}{d\Omega} =\frac{\left|T(s,t)\right|^2}{64\pi^2 s}.$}

$$
T(s,t)=-2is\bar{\psi}_{q_2} \int d\vec{b}_\bot e^{i\Delta
\vec{b}_\bot} \left\{\exp \left[i\chi _0 (b)+i\left(\vec{n}\times
\vec{\sigma}\right)_z\chi _1(b)\right]-1\right\}\psi
_{p_2}.  \eqno(3.14)
 $$
Thus, allowance for the anomalous magnetic moment of the nucleon in the eikonal phase leads to appearance of an additive term
responsible for the spin flip in the scattering process. Integrating in Eq $(3.14)$ with respect to the angular variable $[32]$,
we obtain the amplitude
$$
T (s,t)=\bar{\psi }_{q_2} \left[f_0(s,\Delta)+i\sigma_y f_1
(s,\Delta)\right]\psi _{p_2}, \eqno(3.15)
$$
where $f_0(s, \Delta)$ and $f_1 (s, \Delta)$ describe processes with and without spin flip, respectively, and they are given by
$$
f_{0}(s,\Delta)=-4\pi {\rm ?}s\int _0^\infty\rho d\rho J_0 (\Delta
\rho)\left[e^{i\chi _0} \cos \chi _1 -1\right],  \eqno(3.16)
$$
$$f_{1}(s,\Delta)=4\pi s\int _0^\infty\rho d\rho
J_1(\Delta \rho)\sin \chi _{1}.  \eqno(3.17)$$ \\

It is obvious that all the expressions $(3.14)-(3.17)$ are finite, and therefore the
renormalization problems does not arise in out approximation in the limit $s\rightarrow\infty$.\\

\section{ Coulomb Interference }
Coulomb interference for particles with anomalous magnetic moment was considered for the first time in Ref. $[39]$, in which the amplitude was actually only in the first Born approximation in the Coulomb interaction. The relativistic eikonal approximation was used for the first time to calculate Coulomb interference without allowance for spin $[34]$. It is interesting to use our results to consider Coulomb interference $ [33-39]$ in the scattering of the charges hadrons $\pi N$. The nuclear interaction can be included in our approach by replacing the eikonal phase in accordance with $[34]$ $ \chi_{em}(b) \rightarrow \chi_{em} (b) + \chi_h (b)$

$$
T \left(s,t\right)=-2is\bar{\psi }_{q_{2} } \int
d\vec{b}_{\bot } e^{i\Delta \vec{b}_{\bot } } \left(\exp \left[i\chi
_{em} \left(b\right)+i\chi _{h} \left(b\right)\right]-1\right)\psi
_{p_{2} }, \eqno(4.1)
$$
where $ \chi_{em}(b)= \chi_0 (b)+ i [\overrightarrow{n}\times \overrightarrow{\sigma}]_z \chi_1(b)$,  is eikonal phase that corresponds to the nuclear interaction. For the following discussion, the Eq. $(4.1)$ is rewritten in the form

$$
T(s,t)=T_{em} (s,t)+T_{eh}(s,t), \eqno(4.2)
$$
where $ T_{em}$ is the part of the scattering amplitude due to the electromagnetic interaction and determined by Eq. $(3.14)$ or
formulas $ (3.15)-(3.17)$, and $ T_{eh}(s,t)$ is the interference electromagnetic hadron part of the scattering amplitude
$$
T_{eh}(s,t)=e^{\varphi _t} T_h (s,t)=-2is\bar{\psi}_{q_2} \int
d\vec{b}_\bot e^{i\Delta \vec{b}_\bot}\left(e^{i\chi _h(b)}
-1\right)e^{i\chi _{em} (b)} \psi _{p_2}, \eqno(4.3)
$$
here $ \phi_t $ is the sum of the phase of the Coulomb and nuclear interaction, $T_h(s,t)$ is the purely nuclear amplitude obtained in
the absence of an electromagnetic interaction. In the region of high energies $ s \rightarrow \infty$, $|t|/s \rightarrow 0 $, it is
sufficient to retain only the terms linear  in $ \kappa$  because $ \kappa$  is  small in the all  the following calculations.
Integrating in the expression $(3.15)$ , we obtain
$$
T_{em} (s,t)=\frac{8\pi \alpha s}{\Delta ^2} \frac{\Gamma (1-i\alpha
)}{\Gamma (1+i\alpha)} \exp \varphi _{em} \bar{\psi}_{q_2}
\left[1-i\frac{\kappa}{e}\sigma_y \Delta \right]\psi_{p_2},\quad
\varphi _{em} =ie\left[\ln \frac{\Delta ^2}{\mu ^2} -2\gamma\right],
\eqno(4.4)
$$
where $\alpha= e^2 /4\pi $, $\mu$ is the photon mass, and $\gamma=0,577215...$ is the Euler constant. Calculating $T_{ch}(s,t)$ we
use the standard formulas
$$
T_{h} (s,t)=\bar{\psi }_{q} f_{h} (s,t=0)\psi _{p} e^{R^{2} t}
,\quad t=-\Delta ^{2}\eqno(4.5)
$$
where
$$
f_h(s,t=0)=s\; \sigma _{tot} \left[i+\frac{Ref_h (s,t=0)}{Imf_h(s,t=0)} \right]. \eqno(4.6)
$$
Then, calculating the integral $(4.3)$, we obtain
$$
T_{emh} (s,t)=T_{h} (s,t)\left[1+\frac{e\kappa }{4\pi } \sigma _{y}
\Delta \right]\exp \varphi _{t}, \quad \varphi _{t} =-ie\left[\ln
\left(R\mu \right)^{2} +2\gamma\right].  \eqno(4.7)
$$
Hence, for the difference of the (infinite) pases of the amplitudes $ T_{eh}$ and $T_c (s,t)$ we find the expression
$$
\varphi =\varphi _t-\varphi _c =-i\alpha \ln (R\Delta)^2.  \eqno (4.8)
$$
In contrast to $[39]$ in which Coulomb interference with allowance for anomalous magnetic moment, in our approach we have exactly
summed all ladder and cross- lader Feynman graphs. In the case of scattering through small angles $\Delta=2psin\frac{\theta}{2} \simeq
p\theta $, $p$ is the relativistic momentum in cms), the phase difference is equal to $\phi= 2i\alpha ln \frac{1}{Rp\theta}$. This
result is practically the same as Bethe's $[33]$

\section{Conclusions }

In the framework of the functional integration, a method is proposed  for studying the scattering of a scalar pion on nucleon with an anomalous magnetic moment in quantum field theory. We obtained an eikonal representation of the scattering amplitude in the asymptotic region $s\rightarrow \infty$, $ \mid t \mid \ll s$. Allowance for the anomalous magnetic moment leads to the additional terms in the amplitude that do not vanish as $s\rightarrow \infty$, and these describe spin flips of the particles in the scattering process. It is shown that in the limit $s\rightarrow \infty $ in the eikonal approximation the renormalization problem does not arise since the unrenomalized divergences disappear in this approximation. As an application of the eikonal formula obtained, we considered the Coulomb interference in the scattering of charged hadrons, and we found a formula for the phase difference, which generalizes the Bethe's formula in the framework of relativistic quantum field theory.

{\bf Acknowledgments.} We would like to express gratitude to Profs. B.M. Barbashov, A.V.Efremov, V.N. Pervushin for useful
discussions. N.S.H. is also indebted to Profs. Randjbar-Daemi and GianCarlo Ghirardi for support during my stay at the Abdus Salam ICTP in Trieste.
This work was supported in part by the Abdus Salam ICTP , JINR Dubna, TRIGA and Vietnam National University under Contract QG.TD.10.02.

\vskip 0.5cm

\section*{Appendix: The Green's functions in the form of a functional integral $[40]$}
In this appendix we find the representation of the Green's functions of the Klein-Gordon equation and the Dirac equation for single particles in an external electromagnetic field $A_\mu (x)$, $\partial A_\mu (x)/\partial x_\mu =0$ in the form of a functional integral. Let us consider the Klein-Gordon equation for the Green' function \footnote{Here we use all the notations presented in Ref. [4]}

$$
[(i\partial_{\mu}+eA_{\mu}(x))^2 -m^2] G(x,y|A)=-\delta^{4}(x-y). \eqno(A.1)
$$

 Writing the inversion operator in exponential form, as proposed by Fock $[41]$ and Feynman $[42]$ , we express the solution  of Eq. $(A.1)$  in an operator form

$$
G(x,y|A)=i\int_{0}^{\infty}d\xi\exp\bigg\{i\int_{0}^{s}(i\partial_{\mu}(\xi)+eA_{\mu}(x,\xi))^2-im^2 \bigg\}\delta^{4}(x-y), \eqno(A.2)
$$

the exponent in expression $(A.2)$, which contains the non-commuting operators $\partial_{\mu}(x,\xi)$ and $A_{\mu}(x,\xi)$ is considered as $T_{\xi}$-exponent, where the ordering subscript $\xi$ has meaning of proper time divided by mass $m$. All operators in $(A.2)$ are assumed to be commuting functions that depend on the parameter $\xi$. The exponent in eq. $(A.2)$ is quadratic in the differential operator $ \partial_{\mu}$. However, the transition from $T_{\xi}$-exponent to an ordinary operator expression ("disentangling" the differentiation operators in the argument of the exponential function by terminology of Feynman $[42]$) cannot be performed without the series expansion with respect to an external field. But one can lower the power of the operator  $\partial_{\mu}(x,\xi)$ in eq. $(A.2)$ by using the following formal transformation

$$
\exp\bigg\{i\int_{0}^{s}d\xi (i\partial_{\mu}(\xi)+eA_{\mu}(x,\xi))^2 \bigg\}=C\int\delta^{4}\nu \exp\bigg\{-i\int_{0}^{s}\nu_{\mu}^{2}(\xi)d\xi+2i\int_{0}^{s}\bigg[i\partial_{\mu}(\xi)+eA_{\mu}(x,\xi)\bigg] \bigg\}. \eqno(A.3)
$$

The functional integral in the right-hand side of eq.$(A.3)$ is taken in the space of 4-dimensional function $ \nu_{\mu}(\xi)$ with a Gaussian measure. The constant $C_{\mu}$ is defined by the condition:

$$
C_{\mu}\int\delta^{4}\nu_{\mu} \exp \bigg \{ -i\int \nu_{\mu}^{2}(\xi) d\xi\bigg\}=1. \eqno(A.4)
$$

After substituting $ (A.3)$ into $(A.2)$, the operator $\exp \bigg [2i\int_{0}^{s}\nu^{\mu}(\xi)\partial_{\mu}(\xi)\bigg]$ can be "disentangled" and we can find a solution in the form of the functional integral:

$$
G(x,y|A)=i\int_{0}^{s}ds e^{-im^2 s} \int [\delta^{4}\nu]_{0}^{s}\exp\bigg[ie\int_{0}^{s} 2\nu_{\mu}(\xi)A_{\mu}(x-2\int_{\xi}^{s}\nu(\eta)d\eta) \bigg]\delta^{4}(x-y-2\int_{\xi}^{s}\nu(\eta)d\eta), \eqno(A.5)
$$
where
$$
[\delta^{4}\nu]_{s_1}^{s_2}=\frac{\delta^{4}\exp[-i\int_{s_1}^{s_2} \nu_{\mu}^{2}(\eta)d\eta]\Pi_{\eta}d^{4}\eta}{\int\delta^{4}\exp[-i\int_{s_1}^{s_2} \nu_{\mu}^{2}(\eta)d\eta]\Pi_{\eta}d^{4}\eta},
$$

and $[\delta^{4}\nu]_{s_1}^{s_2}$ is volume element of the functional space of the four-dimensional functions $\nu_{\mu}(\eta)$ defined in the interval $ s_1\leq \eta \leq s_2 $. \\

The expression for the Fourier transform of the Green's function $(A.5)$ takes the form.
$$
G(p,q|A)=\int d^{4}xd^{4}y G(x,y|A)=i\int_{0}^{s}d\xi e^{i(p^2-m^2)s}\int d^{4}x e^{i(p-q)x}\int[\delta^{4}\nu]_{0}^{s}\exp[ie\int_{0}^{s}J_{\mu}A_{\mu}], \eqno(A.6)
$$

here we use the notation $ \int JA=\int J_{\mu}(z)A_{\mu}(z)$, and $J_{\mu}(z)$ is the current of the particle 1 defined by

$$
J_{\mu}(z)=2\int_{0}^{s}\nu_{\mu}(\xi)\delta(z-x_i +2p_i\xi+2\int_{0}^{\xi}\nu_i(\eta)d\eta). \eqno(A.7)
$$

Up to this point, we have found the closed expression for the Green's function of single spinless particles in an external given field in the form of functional integral. In a similar manner we find the representation of the Green's function for the Dirac equation,
$$
[ i\gamma_{\mu}\partial_{\mu}-m+e\gamma_{\mu}A_{\mu}(x)]G(x,y|A)=-\delta^{4}(x-y). \eqno(A.8)
$$
Since functional integrals are related to the solution of second - order equations, it is convenient to introduce the squared Green's function
$ \overline{G}(x,y|A)$

$$
G(x,y|A)=[i\gamma_{\mu} \partial_{\mu} + m + \gamma_{\mu}A_\mu(x)]\overline{G}(x,y|A),   \eqno(A.9)
$$
in which $ \overline{G}(x,y|A)$ satisfies
$$
[ (i\partial_\mu +eA_\mu(x))^2 -m^2 + e\sigma_{\mu\nu} \partial_{\mu}A_\nu(x)]\overline{G}(x,y|A)=-\delta^{4}(x-y). \eqno(A.10)
$$
Comparing equation Equation $(A.2)$ and $(A.9)$, we get to see some term $ \sigma_{\mu\nu}$ related to spin of particle 2 \footnote{The problem of "disentangling" Dirac matrices in the solution of the Dirac equation in an external field was considered by Fradkin $[43]$}
$$
\overline{G}(x,y|A)=i\int_{0}^{s} e^{-im^2 s}T_{\gamma} \int[\delta^{4}\nu]_{0}^{s}\exp \{ie\int_{0}^{s}J_{\mu}A_{\mu}(x)\}\delta^{4}(x-y-2\int_{\xi}^{s}\nu(\eta)d\eta), \eqno(A.11)
$$
where $ T_\gamma$ is the symbol of ordering the $\gamma_{\mu}$ matrices  with respect to the ordering index $\xi$, and $J_{\mu}(z)$ is the current of the particle 2 defined by
$$
J_{\mu}(z)=2\int_{0}^{s} [\nu_{\mu}(\xi)+ \frac{1}{2}\sigma_{\mu\nu}(\xi)i\partial_{\nu}]\delta(z-x_i +2p_i\xi+2\int_{0}^{\xi}\nu_i(\eta)d\eta). \eqno(A.11)
$$
It is important to notice that the solutions of eqs. (A.2) and (A.9) are similar, however, the one of the latter contains one more term related to the spin. Because $\sigma_{\mu\nu}$ depends on $\xi$ as an ordering index, the solution of eq. (A.9) must contain $\gamma_\xi$, therefore, $T_\xi$ remains in eq. (A.11).

\end{document}